\documentclass[conference,a4paper]{IEEEtran}

\usepackage[utf8]{inputenc} 
\usepackage[T1]{fontenc}
\usepackage{url}
\usepackage{ifthen}
\usepackage{cite}
\usepackage[cmex10]{amsmath} 
\usepackage{amssymb}
\usepackage{graphicx}
\usepackage{subfigure}
\usepackage{epstopdf}

\newtheorem{thm}{Theorem}[section]
\newtheorem{lem}[thm]{Lemma}

\newtheorem{cor}[thm]{Corollary}

\newtheorem{exmp}[thm]{Example}

\begin{document}
\title{Optimal Error Correcting Delivery Scheme for Coded Caching with Symmetric Batch Prefetching} 


\author{%
  \IEEEauthorblockN{Nujoom Sageer Karat, Anoop Thomas and B. Sundar Rajan}
  \IEEEauthorblockA{Department of Electrical Communication Engineering, Indian Institute of Science, Bengaluru 560012, KA, India \\
E-mail: \{nujoom,thomas,bsrajan\}@iisc.ac.in}
}

\maketitle

\begin{abstract}
 Coded caching is used to reduce network congestion during peak hours. A single server is connected to a set of users through a bottleneck link, which generally is assumed to be error-free. During non-peak hours, all the users have full access to the files and they fill their local cache with portions of the files available. During delivery phase, each user requests a file and the server delivers coded transmissions to meet the demands taking into consideration their cache contents. In this paper we assume that the shared link is error prone. A new delivery scheme is required to meet the demands of each user even after receiving finite number of transmissions in error. We characterize the minimum average rate and minimum peak rate for this problem. We find closed form expressions of these rates for a particular caching scheme namely \textit{symmetric batch prefetching}. We also propose an optimal error correcting delivery scheme for coded caching problem with symmetric batch prefetching.
\end{abstract}
\section{INTRODUCTION}

Coded caching techniques are aimed at reducing peak hour traffic in networks \cite{MaN}. A portion of the content is made locally available at the users during non-peak periods so that traffic can be reduced at peak hours. The seminal work in \cite{MaN} shows that apart from the \textit{local caching gains} obtained by placing contents at user caches before the demands are revealed, a \textit{global caching gain} can be obtained by coded transmissions. The fundamental scheme developed in \cite{MaN} is a centralized coded caching scheme, where all users are linked to a single fixed server. This scenario is extended to a more realistic decentralized scheme in \cite{MaN2}. More extensions to non-uniform demands \cite{NiM} and online coded caching \cite{PMN} are also available in literature. 

A coded caching scheme has two phases:  a placement phase and a delivery phase. In the placement phase or prefetching phase, each user can fill their local cache memory using the entire database. During this phase there is no bandwidth constraint as the network is not congested and the only constraint here is the memory. Delivery phase is carried out once the users reveal their demands. During the delivery phase only the server has access to the file database and the constraint here is the bandwidth as the network is congested in this phase. During placement phase some parts of files have to be judiciously cached at each user in such a way that the rate of transmission is reduced during the delivery phase. The prefetching can be done with or without coding. If during prefetching, no coding of parts of files is done, the prefetching scheme is referred to as uncoded prefetching \cite{MaN}, \cite{YMA}. If coding is done during prefetching stage, then the prefetching scheme is referred to as coded prefetching \cite{CFL}.

In this work an extension to the coded caching problem called as the error correcting coded caching scheme is considered. In our model, the receivers should be able to decode their demanded messages even when the delivery phase has finite number of transmission errors. The placement phase is assumed to be error-free. This assumption can be justified as during placement phase there is no bandwidth constraint and any number of re-transmissions can be done to make the placement error-free. A similar model in which the delivery phase takes place over a packet erasure broadcast channel was considered in \cite{BWT}.

Error correction at receivers can be achieved by increasing the number of transmissions. This paper addresses the problem of finding error correcting delivery schemes which use minimum number of transmissions. The main contributions of this paper are as follows.
\begin{itemize}
	\item We characterize the minimum average rate and minimum peak rate for error correcting coded caching scheme (Section \ref{sec:sys_model}).
	\item We prove the optimality of a particular delivery scheme introduced in \cite{YMA}, which we refer to as the Yu Maddah-Ali Avestimehr (YMA) scheme, with a particular prefetching scheme
	 namely symmetric batch prefetching, using results from index coding (Section \ref{sec: YMA_opt}).
	\item  For error correcting delivery scheme for coded caching problem with symmetric batch prefetching, we find closed form expressions for expected minimum average rate and minimum peak rate (Section \ref{sec:error_cor}).
	\item We propose an optimal error correcting delivery scheme for coded caching problem with symmetric batch prefetching (Section \ref{sec:error_cor}). 
\end{itemize}

Throughout the paper $\mathbb{F}_{q}$ denotes the finite field with $q$ elements, where $q$ is a power of a prime, and $\mathbb{F}^{*}_{q}$ denotes the  set of all nonzero elements of $\mathbb{F}_q$. The notation $[K]$ is used for the set $\{1,2,...,K\}$ for any integer $K$. For a $K \times N$ matrix $L$, $L_i$ denotes its $i$th row. For a set $S \subseteq [K]$, $L_S$ denotes the $ |S| \times N$ matrix obtained from $L$ by deleting the rows of $L$ which are not indexed by the elements of $S$. We denote $\bf{e_\textit{i}}$ $= (\underbrace{0,...,0}_{i-1},1,\underbrace{0,...,0}_{K-i}) \in \mathbb{F}^{n}_{q}$ as the unit vector having a one at the $i$th position and zeros elsewhere. A linear $[n,k,d]_q$ code $\mathcal{C}$ over $\mathbb{F}_q$ is a $k$-dimensional subspace of $\mathbb{F}^{n}_{q}$ with minimum Hamming distance $d$. The vectors in $\mathcal{C}$ are called codewords. A matrix ${G}$ of size $k \times n$ whose rows are linearly independent codewords of $\mathcal{C}$ is called a generator matrix of $\mathcal{C}$. A linear $[n,k,d]_q$ code $\mathcal{C}$ can thus be represented using its generator matrix ${G}$ as,
$ \mathcal{C} = \{ {\bf{y}}G: {\bf{y}} \in \mathbb{F}^{k}_{q} \}.$ We let $N_{q}[k,d]$ denote the length of the shortest linear code over $\mathbb{F}_q$ which has dimension $k$ and minimum distance $d$.

\section{Preliminaries and Background}
In this section we revisit a few results from error correcting index coding problems \cite{DSC} which are used in this paper and discuss the symmetric batch prefetching scheme and YMA delivery scheme \cite{MaN} \cite{YMA}. We also review the connection between index coding and coded caching problems \cite{MaN}. 

\subsection{Index Coding Problem}
The index coding problem with side information was introduced in \cite{BiK}. A single source has $n$ messages $x_1,x_2 \ldots , x_n$ where $x_i \in \mathbb{F}_{q}, ~\forall i \in [n].$ There are $K$ receivers, $R_1, R_2 \ldots, R_K$. Each receiver possesses a subset of messages as side information. Let $\mathcal{X}_i$ denote the set of  indices of the messages belonging to the side information of receiver $R_i$. The map $f:[K] \rightarrow [n] $ assigns receivers to indices of messages demanded by them. Receiver $R_i$ demands the messages $x_{f(i)}$, $f(i) \notin \mathcal{X}_i$ \cite{DSC}. The source knows the side information available to each receiver and has to satisfy the demand of each receiver in minimum number of transmissions. An instance of index coding problem can be completely characterized by a side information hypergraph \cite{AHLSW}. Given an instance of the index coding problem, finding the best \textit{scalar linear} binary index code is equivalent to finding the \textit{min-rank} of the side information hypergraph \cite{DSC}, which is known to be an NP-hard problem in general \cite{BBJ}, \cite{RP}.

%

An index coding problem with $K$ receivers and $n$ messages can be represented by a hypergraph $\mathcal{H} (V,E)$, where $V=[n]$ is the set of vertices and $E$ is the set of hyperedges \cite{AHLSW}. Vertex $i$ represents the message $x_i$ and each hyperedge represents a receiver. In \cite{DSC}, the min-rank of a hypergraph $\mathcal{H}$ over $\mathbb{F}_q$ is defined as,
\begin{equation*}
\kappa_q(\mathcal{H}) \triangleq
\min\{\text{rank}_q(\{{\bf{v_i}}
+{\bf{e_i}}\}_{i\in [K]}): \\
{\bf{v_i}} \in \mathbb{F}^{n}_q, {\bf{v_i}} \triangleleft \mathcal{X}_i\},
\end{equation*}
where $\bf{v_i}$ $\triangleleft$ $\mathcal{X}_i$ denotes that $\bf{v_i}$ is the subset of the  support of $\mathcal{X}_i$; the support of a vector $\bf{u}$ $\in \mathbb{F}^{n}_{q}$ is defined to be the set $\{i\in [n]: u_i \neq 0  \}$. This min-rank defined above is the smallest length of scalar linear index code for the problem. A linear index code of length $N$ can be expressed as $XL$, where $L$ is an $n \times N$ matrix and $X = [x_1~x_2 \ldots ~x_n]$. The matrix $L$ is said to be the \textit{matrix corresponding to the index code}. 

Let $\mathcal{G}$ = $(\mathcal{V},\mathcal{E})$ be an undirected graph, then a subset of vertices $\mathcal{S}$ $\subseteq$ $\mathcal{V}$ is called an independent set if $\forall u, v \in \mathcal{S}$, $\{u,v\}$ $\notin$ $\mathcal{E}$. The size of a largest independent set in the graph $\mathcal{G}$ is called the independence number of $\mathcal{G}$. Dau {\it{et al}}. in \cite{DSC} extended the notion of independence number to the case of directed hypergraph corresponding to an index coding problem. For each receiver $R_i$, define the sets $$
\mathcal{Y}_i \triangleq [n] \setminus \bigg( \{f(i) \} \cup \mathcal{X}_i \bigg) $$
and
$$\mathcal{J(\mathcal{H})} \triangleq \cup_{i\in [K]} \{\{f(i)\} \cup Y_{i} : Y_i \subseteq \mathcal{Y}_i\}.$$
A subset $H$ of $[n]$ is called a generalized independent set in $\mathcal{H}$, if every nonempty subset of $H$ belongs to $\mathcal{J(\mathcal{H})}$. The size of the largest independent set in $\mathcal{H}$ is called the generalized independence number and is denoted by $\alpha (\mathcal{H})$.

It is proved in \cite{DSC} that a matrix $L$ corresponds to an index code if and only if
$$ \text{wt} \bigg(\sum_{i \in H} z_iL_i\bigg) \geq 1 $$
for all $H \in \mathcal{J(\mathcal{H})}$ and for all choices of $z_i \in F^*_q$, $i \in H.$

\par The quantities  $\alpha (\mathcal{H})$ and $\kappa_q(\mathcal{H})$ decide the bounds on the optimal length of error correcting index codes. The error correcting index coding problem  with side information was defined in \cite{DSC}. An index code is said to correct $\delta$ errors if after receiving at most $\delta$ transmissions in error, each receiver is able to decode its demand. A $\delta$-error correcting index code is represented as $(\delta, \mathcal{H})$-ECIC. An optimal linear $(\delta, \mathcal{H})$-ECIC over $\mathbb{F}_q$ is a linear $(\delta, \mathcal{H})$-ECIC over $\mathbb{F}_q$ of the smallest possible length $\mathcal{N}_{q}[\mathcal{H},\delta]$. Bounds were established in \cite{DSC} on this length, the lower bound, known as the $\alpha$-bound, is given by
\begin{equation*}
\mathcal{N}_{q}[\mathcal{H},\delta] \geq N_q[\alpha(\mathcal{H}), 2\delta + 1].
\end{equation*}
The upper bound or the $\kappa$-bound is given by
\begin{equation*}
\mathcal{N}_{q}[\mathcal{H},\delta] \leq N_q[\kappa_q(\mathcal{H}), 2\delta + 1].
\end{equation*}
Thus the length of an optimal linear $(\delta,\mathcal{H})$-ECIC over $\mathbb{F}_q$ satisfies
\begin{equation}
\underbrace{N_q[\alpha(\mathcal{H}), 2\delta + 1]~ \leq ~}_{\alpha\text{-bound}}  \mathcal{N}_{q}[\mathcal{H},\delta] \underbrace{~\leq~ N_q[\kappa_q(\mathcal{H}), 2\delta + 1]}_{\kappa\text{-bound}}. \label{eq:bds}
\end{equation}
The $\kappa$-bound is achieved by concatenating an optimal linear classical error correcting code and an optimal linear index code. Thus for any index coding problem, if $\alpha (\mathcal{H})$ is same as $\kappa_q(\mathcal{H})$, then concatenation scheme would give optimal error correcting index codes \cite{SaR}, \cite{SagR}, \cite{SSR}. 

\subsection{Symmetric batch prefetching and YMA scheme}
In this paper we consider one particular type of uncoded prefetching scheme referred to as \textit{symmetric batch prefetching} \cite{MaN}, \cite{YMA}. We denote this prefetching scheme as $\mathcal{M}_{SB}$. There are $K$ users and the server has $N$ files $X_1, X_2, \ldots ,X_N$, each of $F$ bits. Each user has a cache of size $MF$ bits.  The demand vector is denoted by $\textbf{d} = (d_1, ..., d_K)$, where $d_i$ is the index of the file demanded by user $i$. The number of distinct files requested in $\textbf{d}$ is denoted by $N_e(\textbf{d})$. Each file is partitioned into $ K \choose r$ non-overlapping subfiles of approximately equal sizes, where $r$ is a parameter defined as $r=\frac{KM}{N}$. Each subfile of file $i$ is denoted as $X_{i, \mathcal{A}}$, where $\mathcal{A} \subseteq \{1, ..., K\}, \vert \mathcal{A} \vert = r$. Now, each user $k \in \{1,2,\ldots,K \}$ caches all subfiles $X_{i, \mathcal{A}}$ of the message $X_i$, which satisfy the criterion  $k \in \mathcal{A}$. Since each user caches ${{K-1} \choose {r-1}}N$ subfiles and each subfile has $F/{K \choose r}$ bits, the number of bits cached at each user equals $NrF/K = MF$ bits. Thus all the available cache memory for the users is fully utilized. The delivery scheme proposed in \cite{YMA} is optimal for symmetric batch prefetching. We refer to this scheme as YMA scheme. In \cite{YMA} it has been shown that for YMA scheme the minimum average rate $R^*$ for $ r \in \{0,1, \ldots , K\}$ is  
$$ R^*= \mathbb{E}_{\textbf{d}}\bigg[\frac{{K \choose r+1 } - {K-N_e(\textbf{d}) \choose r+1}}{{K \choose r}}\bigg].$$
Furthermore, for $ r\notin \{0,1, \ldots , K\}$, $R^*$ equals the lower convex envelope of its values at $ r \in \{0,1, \ldots , K\}$. The minimum peak rate $R^*_{\text{worst}}$ for $ r \in \{0,1, \ldots , K\}$ is
$$ R^*_{\text{worst}} = \frac{{K \choose r+1 } - {K-\min\{K,N\} \choose r+1}}{{K \choose r}}.$$
Furthermore, for $ r\notin \{0,1, \ldots , K\}$,  $R^*_{\text{worst}}$ equals the lower convex envelope of its values at $ r \in \{0,1, \ldots , K\}$. 
\subsection{Equivalent Index Coding Problems of a Coded Caching Problem}
For a fixed prefetching $\mathcal{M}$ and for a fixed demand \textbf{d}, the delivery phase of a coded caching problem is an index coding problem \cite{MaN}. In fact, for fixed prefetching, a coded caching scheme consists of $N^K$  parallel index coding problems one for each of the $N^K$ possible user demands. Thus finding the minimum achievable rate for a given demand $\textbf{d}$ is equivalent to finding the min-rank of the equivalent index coding problem induced by the demand $\textbf{d}$.
\label{sec:index_caching}

\section{SYSTEM MODEL}
In this section we describe the system model for error correcting coded caching scheme. An error correcting coded caching scheme has a server connected to $K$ users through a shared link which is error prone. 
The server has access to $N$ files $X_1, X_2, ..., X_N$, each of size $F$ bits. Every user has an isolated cache with memory $MF$ bits, where $M \in [0,N]$.  A prefetching scheme is denoted by ${\mathcal{M}}$. 

During the delivery phase, only the server has access to the database. Every user demands one of the $N$ files. The demand vector is denoted by $\textbf{d} = (d_1, ..., d_K)$, where $d_i$ is the index of the file demanded by user $i$. The number of distinct files requested in $\textbf{d}$ is denoted by $N_e(\textbf{d})$. The set of all possible demands is denoted by $\mathcal{D} = \{1, ..., N\}^K.$ During the delivery phase, the server informed of the demand $\textbf{d}$, transmits a signal $Y$, which is a function of $X_1, ..., X_N$, over a shared link. Using the values of bits $\mathcal{M}_k$ and the signal $Y$ each user $k$ needs to reconstruct the requested file $X_{d_{k}}$ even after receiving $\delta$ transmissions in error. 

For the $\delta$-error correcting coded caching problem we define that a communication rate $R(\delta)$ is \textit{achievable} for demand $\textbf{d}$ if and only if there exists a transmission signal $Y$ of $R(\delta)F$ bits such that every user $k$ is able to recover its desired file $X_{d_{k}}$ even after at most $\delta$ transmissions are in error. We denote $R^*(\textbf{d}, \mathcal{M}, \delta)$ as the minimum achievable rate for a given $\textbf{d}$, $\mathcal{M}$ and $\delta$. We define the average rate $R^*(\mathcal{M}, \delta)$ as the expected minimum average rate given $\mathcal{M}$ and $\delta$ under uniformly random demand. Thus $$ R^*(\mathcal{M}, \delta) = \mathbb{E}_{\textbf{d}}[R^*(\textbf{d}, \mathcal{M}, \delta)].$$

The average rate depends on the prefetching scheme $\mathcal{M}$. The minimum average rate  $ R^*(\delta)= \min_{\mathcal{M}} R^*(\mathcal{M}, \delta)$  is the minimum rate of the delivery scheme over all possible $\mathcal{M}$. The rate-memory trade-off for average rate is finding the minimum average rate $R^*(\delta)$ for different memory constraints $M$. 

Another quantity of interest is the peak rate, denoted by $R^*_{\text{worst}}(\mathcal{M}, \delta)$, which is defined as
$$R^*_{\text{worst}}(\mathcal{M}, \delta) = \max_{\textbf{d}} R^*(\textbf{d}, \mathcal{M}, \delta).$$ 
The minimum peak rate is defined as
$$ R^*_{\text{worst}}(\delta)= \min_{\mathcal{M}} R^*_{\text{worst}}(\mathcal{M}, \delta).$$

We denote an index coding problem induced from a coded caching problem for a given demand $\textbf{d}$ as $\mathcal{I}(\mathcal{M}, \textbf{d})$. The side information hypergraph for this index coding problem is denoted by $\mathcal{H}(\mathcal{M}, \textbf{d}).$ The generalized independence number and min-rank of this index coding problem are denoted by $\alpha({\mathcal{M}, \textbf{d}})$ and $\kappa({\mathcal{M}, \textbf{d}})$ respectively.   
A solution to this index coding problem  is thus clearly a solution for the coded caching problem with a given prefetching $\mathcal{M}$ and a given demand $\textbf{d}$. Thus finding $R^*(\textbf{d}, \mathcal{M}, \delta=0)$ is equivalent to finding the min-rank of the induced index coding problem $\mathcal{I}(\mathcal{M}, \textbf{d})$. Extending the argument a little further, we can say that finding $R^*(\textbf{d}, \mathcal{M}, \delta)$ is equivalent to finding the optimal length of $\delta$-error correcting index code for the induced index coding problem $\mathcal{I}(\mathcal{M}, \textbf{d}).$ 

Consider the index coding problem $\mathcal{I}(\mathcal{M}_{SB}, \textbf{d})$ induced from symmetric batch prefetching and for a given demand $\textbf{d}$. Each subfile $X_{i,\{a_1, ..., a_r\}}$ is taken as a message of the index coding problem. Hence there are a total of $N{K \choose r}$ messages. Each user now can be split into ${K \choose r}$ receivers each demanding one message. Thus in total there are $K{K \choose r}$ receivers for the induced index coding problem. The cache content can be taken as side information. Thus each receiver now has $N{K-1 \choose r-1}$ messages as side information. Consider an index code of length $\ell$ for $\mathcal{I}(\mathcal{M}_{SB}, \textbf{d})$. This means that using $\ell$ transmissions, all the receivers decode their demands. Each transmission here is of $F/{K \choose r}$ bits. Thus the total number of bits transmitted is $\ell F/{K \choose r}$. Thus an achievable rate $R(\delta=0)$ for this coded caching scheme is $\ell /{K \choose r}$.   

In the subsequent sections we prove that $\alpha({\mathcal{M}_{SB}, \textbf{d}}) = \kappa({\mathcal{M}_{SB}, \textbf{d}})$ for all $\textbf{d}$ of the coded caching problem with symmetric batch prefetching. It is known that YMA scheme is optimal for this prefetching \cite{YMA}. Since $\alpha({\mathcal{M}_{SB}, \textbf{d}}) = \kappa({\mathcal{M}_{SB}, \textbf{d}})$, $\alpha$ and $\kappa$ bounds for the optimum length for $\delta$-error correcting index codes are also equal (\ref{eq:bds}). Hence the concatenation scheme of the optimal index code and the optimal error correcting code is optimal. Thus the optimal delivery scheme for the $\delta$-error correcting coded caching scheme with symmetric batch prefetching is to concatenate YMA scheme with optimal classical error correcting code.
\label{sec:sys_model}

\section{Optimality of YMA Scheme with symmetric batch prefetching}
In \cite{YMA} it has been proved that YMA scheme is optimal for symmetric batch prefetching. In this section, we present an alternative and simpler proof for  the optimality of YMA scheme using results from index coding. Moreover, the results presented in this section are used to construct optimal error correcting delivery scheme for coded caching problem with symmetric batch prefetching as shown in Section \ref{sec:error_cor}. 

The lemma below gives a lower bound for $\alpha({\mathcal{M}_{SB}}, \textbf{d})$ which is used to prove the optimality of YMA scheme.

\begin{lem}
For the index coding problem $\mathcal{I}(\mathcal{M}_{SB}, \textbf{d})$ corresponding to a coded caching problem with symmetric batch prefetching and demand $\textbf{d}$, $$\alpha({\mathcal{M}_{SB}, \textbf{d}}) \geq {K \choose r+1 } - {K-N_e(\textbf{d}) \choose r+1},$$ for $ r \in \{0, 1, ..., K\}.$  \label{lem:sb_alpha}
\end{lem}

\begin{IEEEproof}
Recall that $N_e(\textbf{d})$ is the number of distinct files requested in $\textbf{d}$ . Without loss of generality we can assume that the first $N_e(\textbf{d})$ receivers have distinct demands. We construct a set $B$, whose elements are messages of the index coding problem $\mathcal{I}(\mathcal{M}_{SB}, \textbf{d})$ such that the set of indices of the messages in $B$ forms a generalized independent set. The set $B$ is constructed as $$  B= \bigcup_{ i \in [N_e(\textbf{d})]} \{X_{d_i, \{a_1, ..., a_r\}}: a_1, ..., a_r \neq d_1, d_2, ...,d_i\}.$$
	 Let $H$ be the set of indices of the messages in $B$. The claim is that  $H$ is a generalized independent set. Each message in $B$ is demanded by one receiver. Hence all the subsets of $H$ of size one are present in $\mathcal{J}(\mathcal{H})$, where $\mathcal{H}$ here is the side information hypergraph corresponding to $\mathcal{I}(\mathcal{M}_{SB}, \textbf{d})$.
	Consider any set $C = \{X_{i1, \{a_{11}, \ldots, a_{r1}\}}, \ldots, X_{ik, \{a_{1k}, \ldots, a_{rk}\}}  \} \subseteq B$ where $i1 \leq i2 \leq \ldots \leq ik$. Consider the message $X_{i1, \{a_{11}, \ldots, a_{r1}\}}$. The receiver demanding this message does not have any other message in $C$ as side information. Thus indices of messages in $C$ lie in $\mathcal{J}(\mathcal{H})$. Thus any subset of $H$ lies in $\mathcal{J}(\mathcal{H})$. 	
	Since $H$ is a generalized independent set, we have, $\alpha({\mathcal{M}_{SB}, \textbf{d}}) \geq |H| $. Note that $|H|=|B|$. Number of messages of the form $X_{i, \{a_1, ..., a_r\}}$ which are present in $B$ is ${K-i \choose r}$. Thus $$|B| = \sum_{i=1}^{N_e(\textbf{d})} {K-i \choose r} = \sum_{i=1}^{K} {K-i \choose r} - \sum_{N_e(\textbf{d})+1}^{K} {K-i \choose r}.$$
	To simplify these summations we use the \textit{hockey stick identity} \cite{HS} $\big(\sum_{i=0}^{K}{i \choose r}= {K+1 \choose r+1}\big)$ and make appropriate substitutions. Then we get
	$$ |B|=|H| =  {K \choose r+1 } - {K-N_e(\textbf{d}) \choose r+1},$$
	from which the statement of the lemma follows.
\end{IEEEproof}
Examples to illustrate the construction of generalized independent set and $\alpha({\mathcal{M}_{SB}, \textbf{d}})$ for the induced index coding problem corresponding to a coded caching problem with symmetric batch prefetching are shown below.

\begin{exmp}
	Consider a caching system with $N=K=3$ and $r=M=1$. Since we consider symmetric batch prefetching, the subfiles of each file can be represented as:
	$$ X_1 : X_{1,\{1\}}, ~X_{1,\{2\}}, ~X_{1,\{3\}},$$ $$ X_2: X_{2,\{1\}}, ~ X_{2,\{2\}},~ X_{2,\{3\}} \text{ and }$$ $$
	 X_3: X_{3,\{1\}}, X_{3,\{2\}}, X_{3,\{3\}}.$$
The cache contents are as follows:
$$Z_1: X_{1,\{1\}}, X_{2,\{1\}}, X_{3,\{1\}}, $$
	$$ Z_2: X_{1,\{2\}}, X_{2,\{2\}}, X_{3,\{2\}} \text { and } $$
	$$ Z_3: X_{1,\{3\}}, X_{2,\{3\}}, X_{3,\{3\}}. $$
	\begin{figure} 
		\centering
		(a){%
			\includegraphics[width=0.70\linewidth]{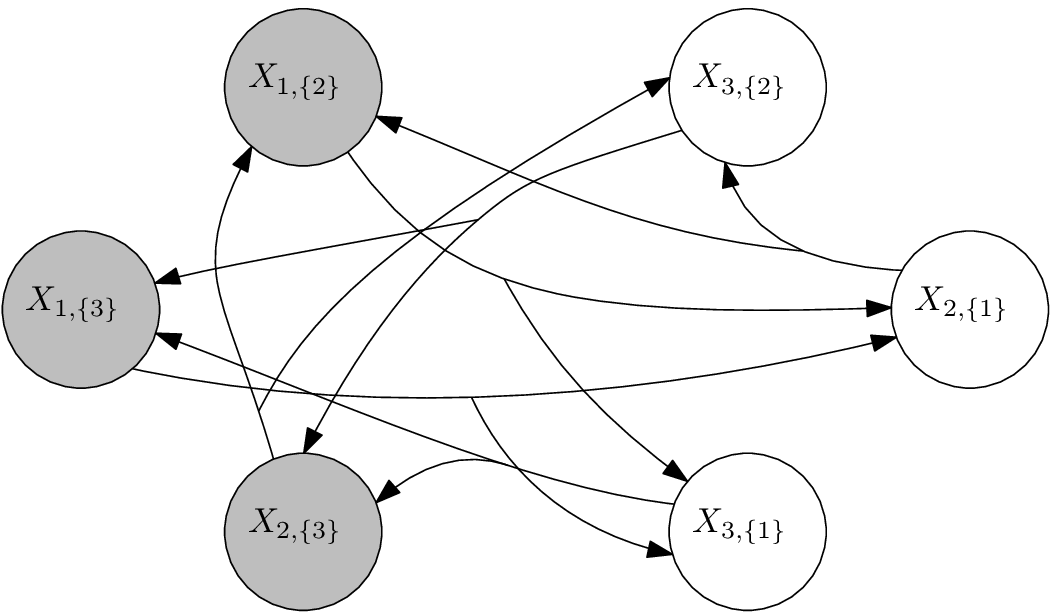}}
		\label{1a} \\ 
		(b){%
			\includegraphics[width=0.55\linewidth]{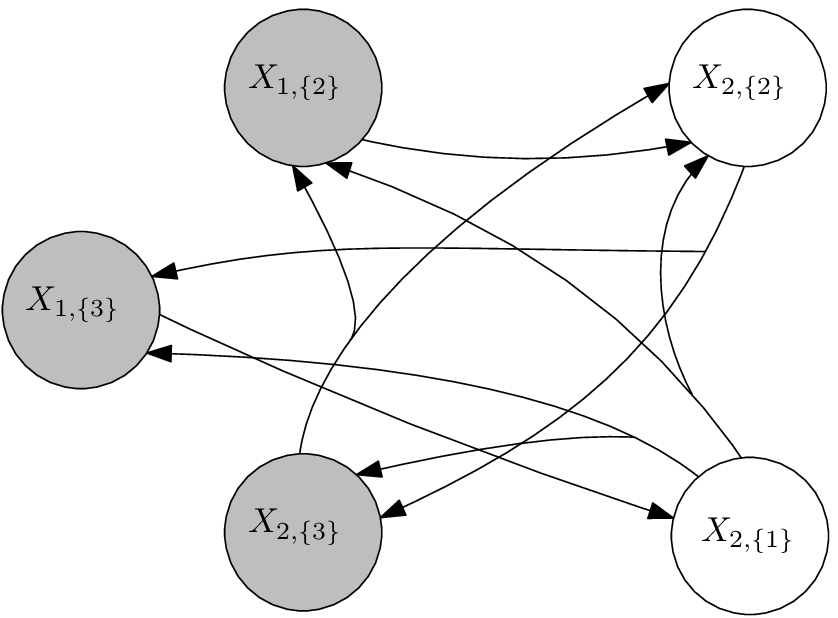}}
		\label{1b}\\
		(c){%
			\includegraphics[width=0.55\linewidth]{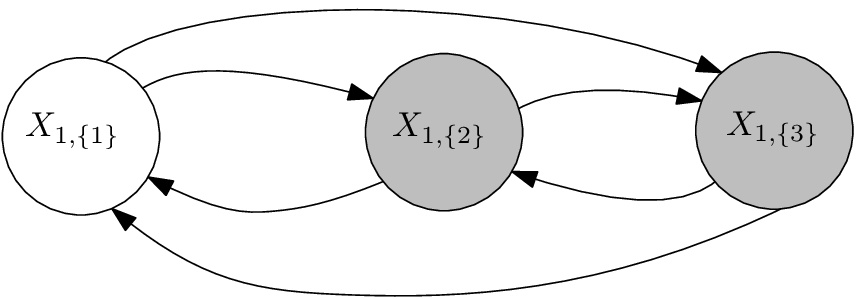}}
		\label{1c}\\
		
		\caption{Side information hypergraphs of the induced index coding problem in Example \ref{Ex:alpha1} with (a) $\textbf{d}=(1,2,3)$ , (b) $\textbf{d}=(1,2,2)$ and (c) $\textbf{d}=(1,1,1)$. The vertices in gray show the elements of $B$ in each case}
		\label{fig1} 
	\end{figure}

Consider $N_e(\textbf{d})=3$ and let the first, the second and the third user demand $X_1$, $X_2$ and $X_3$ respectively. This means $\textbf{d}=(1,2,3)$.
The set $B$ for this case is
	$$ B= \{X_{1,\{2\}}, X_{1,\{3\}}, X_{2,\{3\}}\}.$$
	The side information hypergraph of the induced index coding problem is shown in Fig. \ref{fig1}(a). Note that the messages corresponding to the subfiles which are not demanded by any user are not present in the hypergraph. The vertices in gray show the elements of $B$.

For  $N_e(\textbf{d})=2$ and $\textbf{d} =(1,2,2)$, we get $B=\{X_{1,\{2\}}, X_{1,\{3\}}, X_{2,\{3\}} \}$. The side information hypergraph of the induced index coding problem for this scenario is shown in Fig. \ref{fig1}(b). 

For  $N_e(\textbf{d})=1$ and $\textbf{d} =(1,1,1)$, we get $B=\{X_{1,\{2\}}, X_{1,\{3\}} \}$. The side information hypergraph of the induced index coding problem for this case is shown in Fig. \ref{fig1}(c). 
		
		  \label{Ex:alpha1}
\end{exmp} 

\begin{exmp}
	Consider a caching system with $N=K=3$ and $r=M=2$. Since we consider symmetric batch prefetching, the subfiles are:
	$$ X_1 : X_{1,\{1,2\}}, ~X_{1,\{2,3\}}, ~X_{1,\{1,3\}},$$ $$ X_2: X_{2,\{1,2\}}, ~ X_{2,\{2,3\}},~ X_{2,\{1,3\}} \text{ and }$$ $$
	 X_3: X_{3,\{1,2\}}, X_{3,\{2,3\}}, X_{3,\{1,3\}}.$$
The cache contents according to symmetric batch prefetching are 
$$ Z_1:  X_{1,\{1,2\}}, ~ X_{1,\{1,3\}},  ~X_{2,\{1,2\}}, ~ X_{2,\{1,3\}}, ~ X_{3,\{1,2\}}, ~  X_{3,\{1,3\}} $$
$$ Z_2:  X_{1,\{1,2\}}, ~ X_{1,\{2,3\}},  ~X_{2,\{1,2\}}, ~ X_{2,\{2,3\}}, ~ X_{3,\{1,2\}}, ~  X_{3,\{2,3\}} $$
$$ Z_3:  X_{1,\{2,3\}}, ~ X_{1,\{1,3\}},  ~X_{2,\{2,3\}}, ~ X_{2,\{1,3\}}, ~ X_{3,\{2,3\}}, ~  X_{3,\{1,3\}}. $$	 

Consider $N_e(\textbf{d})=3$ and $\textbf{d}=(1,2,3)$.	 
The set $B$ for this case is
	$$ B= \{X_{1,\{2,3\}}\}.$$
The side information hypergraph of the induced index coding problem and some other index coding problems corresponding to different demands are shown in Fig. \ref{fig2}.

\begin{figure} 
	\centering
	(a){%
		\includegraphics[width=0.40\linewidth]{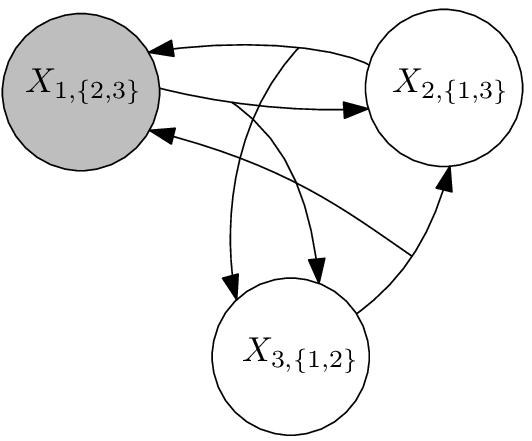}}
	\label{1a} \\ 
	(b){%
		\includegraphics[width=0.40\linewidth]{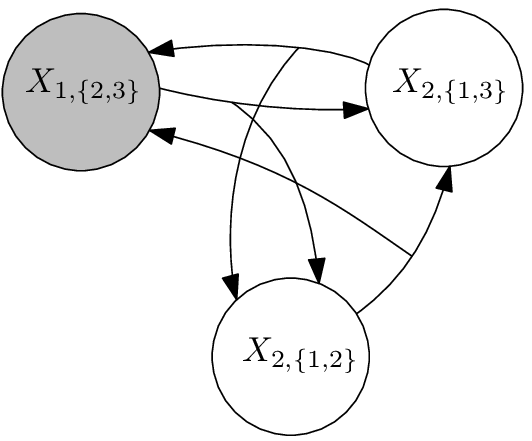}}
	\label{1b}\\
	(c){%
		\includegraphics[width=0.40\linewidth]{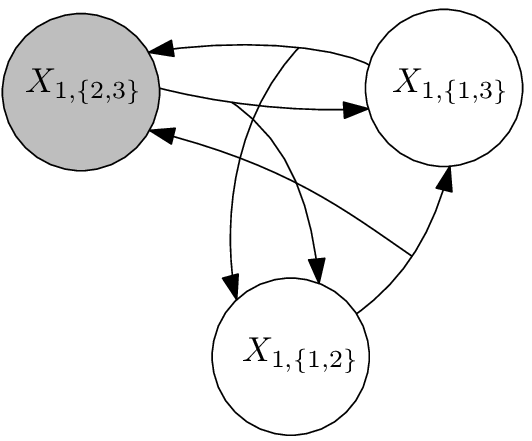}}
	\label{1c}\\
	
	\caption{Side information hypergraphs of the induced index coding problem in Example \ref{Ex:alpha2} with (a) $\textbf{d}=(1,2,3)$ , (b) $\textbf{d}=(1,2,2)$ and (c) $\textbf{d}=(1,1,1)$. The vertices in gray show the elements of $B$ in each case}
	\label{fig2} 
\end{figure}

\label{Ex:alpha2}
		 
\end{exmp} 

The lemma below establishes that $\alpha (\mathcal{H})$ is a lower bound for  $\kappa_q(\mathcal{H})$.  

\begin{lem}
	For an index coding problem with $n$ messages and $K$ receivers represented by a side information hypergraph $\mathcal{H}$, $\alpha (\mathcal{H}) \leq \kappa_q(\mathcal{H}).$ \label{lem:alpha_kappa}
\end{lem} 
\begin{IEEEproof}
	Consider a generalized independent set $S$. Let $L$ be an $n \times \kappa_q(\mathcal{H})$ matrix corresponding to the optimal scalar linear index code for this problem. Since $S \in \mathcal{J}(\mathcal{H})$, we have $\text{wt} \bigg(\sum_{i \in S} z_iL_i\bigg) \geq 1$ for all choices of $z_i \in F^*_q$. By definition, every nonempty subset of $S$ also belongs to $\mathcal{J}(\mathcal{H})$. Thus for any nonempty subset $S'$ of $S$, we have  $\text{wt} \bigg(\sum_{i \in S'} z_iL_i\bigg) \geq 1$. This means that no linear combination of rows of $L_S$ gives an all-zero vector. Thus $\text{rank}(L_S) = |S|$. Therefore, the number of columns, $\kappa_q(\mathcal{H})$, has to be at least $|S|$. From the fact that $|S| \leq \alpha (\mathcal{H})$, the statement of the lemma follows.	
\end{IEEEproof}

\begin{thm}
	YMA scheme proposed in \cite{YMA} is optimal for symmetric batch prefetching. \label{Thm: YMA}
\end{thm}
\begin{IEEEproof}
In \cite{YMA} it has been established that the rate of the YMA scheme for a given demand $\textbf{d}$ with symmetric batch prefetching is $ R^*(\textbf{d}, \mathcal{M}_{SB}, \delta =0) = \frac{{K \choose r+1 } - {K-N_e(\textbf{d}) \choose r+1}}{{K \choose r}},$ for $ r \in \{0,1, \ldots , K\}$. 
If we consider the YMA delivery scheme as the index coded transmission scheme, then the number of transmissions required is ${K \choose r+1 } - {K-N_e(\textbf{d}) \choose r+1}$. Thus the min-rank of the index coding problem is upper bounded as $\kappa({\mathcal{M}_{SB}, \textbf{d}}) \leq {K \choose r+1 } - {K-N_e(\textbf{d}) \choose r+1}.$ Now from Lemma \ref{lem:sb_alpha} and Lemma \ref{lem:alpha_kappa} we get that  $\kappa({\mathcal{M}_{SB}, \textbf{d}}) = {K \choose r+1 } - {K-N_e(\textbf{d}) \choose r+1}.$ 
Hence the statement of the theorem follows.
\end{IEEEproof}
\label{sec: YMA_opt}
\section{Optimal Error Correcting Delivery Scheme for Symmetric Batch Prefetching}
 In this section we give the expression for the average rate and worst case rate for a $\delta$-error correcting delivery scheme for symmetric batch prefetching. Also we propose a $\delta$-error correcting delivery scheme for this case.
 
 \begin{thm}
 	For a coded caching problem with symmetric batch prefetching,
 	$$ R^*(\mathcal{M}_{SB}, \delta) = \mathbb{E}_{\textbf{d}}\bigg[\frac{N_q[\kappa({\mathcal{M}_{SB}, \textbf{d}}), 2\delta+1]}{{K \choose r}}\bigg],$$
 	where $ \kappa({\mathcal{M}_{SB}, \textbf{d}}) = {K \choose r+1 } - {K-N_e(\textbf{d}) \choose r+1},$ for $ r \in \{0,1, \ldots , K\}$. 
 	Furthermore, for $ r\notin \{0,1, \ldots , K\}$, $R^*(\mathcal{M}_{SB}, \delta)$ equals the lower convex envelope of its values at $ r \in \{0,1, \ldots , K\}$. 
 \end{thm}
\begin{IEEEproof}
From Lemma \ref{lem:sb_alpha} and Theorem \ref{Thm: YMA}, we get that for any index coding problem induced from coded caching problem with symmetric batch prefetching, $\alpha({\mathcal{M}_{SB}, \textbf{d}})= \kappa({\mathcal{M}_{SB}, \textbf{d}})$. Thus by (\ref{eq:bds}), the $\alpha$ and $\kappa$ bounds become equal for such index coding problems. The optimal length or equivalently the optimal number of transmissions required for $\delta$ error corrections in those index coding problems is thus $N_q[\kappa({\mathcal{M}_{SB}, \textbf{d}}), 2\delta+1]$ and hence the statement of the theorem follows for $r \in \{0,1, \ldots , K\}$. For  $ r\notin \{0,1, \ldots , K\}$, the lower convex envelope of values of $R^*(\mathcal{M}_{SB}, \delta)$ is achieved by using memory sharing.
\end{IEEEproof} 

\begin{cor}
	For a coded caching problem with symmetric batch prefetching,
$$ R^*_{\text{worst}}(\mathcal{M}_{SB}, \delta) = \frac{N_q[\kappa_{\{\text{worst}\}}({\mathcal{M}_{SB}, \textbf{d}}), 2\delta+1]}{{K \choose r}},$$
	where $ \kappa_{\{\text{worst}\}}({\mathcal{M}_{SB}, \textbf{d}}) = {K \choose r+1 } - {K-\min\{K,N\} \choose r+1}$, for $ r \in \{0,1, \ldots , K\}$. 	Furthermore, for $ r\notin \{0,1, \ldots , K\}$, $R^*_{\text{worst}}(\mathcal{M}_{SB}, \delta)$ equals the lower convex envelope of its values at $ r \in \{0,1, \ldots , K\}$.  	
\end{cor}
\begin{IEEEproof}
	Worst case rate is required when the number of distinct demands is maximum. This happens when $N_e({\textbf{d}})= \min\{K,N\}.$
\end{IEEEproof}
Since the $\alpha$ and $\kappa$ bounds become equal for the induced index coding problems of symmetric batch prefetching, the optimal coded caching delivery scheme here would be the concatenation of the YMA scheme with optimal classical error correcting scheme which corrects $\delta$ errors. Decoding can be done by syndrome decoding for error correcting index codes proposed in \cite{DSC}.

We give a few examples for which we construct optimal error correcting delivery scheme for coded caching problems with symmetric batch prefetching.

\begin{exmp}
Consider the caching system considered in Example \ref{Ex:alpha1} with $N_e(\textbf{d})=3$ and $\textbf{d}=(1,2,3)$. The side information hypergraph of the induced index coding problem is shown in Fig. \ref{fig1}(a). 
When we use YMA scheme, the three transmissions used are:  $ X_{1,\{2\}} \oplus X_{2,\{1\}}, ~X_{2,\{3\}} \oplus X_{3,\{2\}} \text{ and } X_{1,\{3\}} \oplus X_{3,\{1\}}.$ Now if we need to correct $\delta=1$ error, we need to concatenate YMA scheme with a classical error correcting code with optimal length. We have from \cite{Gra}, $N_2[3,3]=6$. One such code is given by the following generator matrix 
	$$
	\bf{G}=
	\begin{bmatrix}
	1 & 0 & 0 & 1 & 1 & 0 \\
	0 & 1 & 0 & 1 & 0 & 1 \\
	0 & 0 & 1 & 0 & 1 & 1
	\end{bmatrix}.
	$$
	 If we concatenate this $[6,3,3]_2$ binary linear code with YMA scheme, we obtain the following six transmissions: $$ Y_1: X_{1,\{2\}} \oplus X_{2,\{1\}},$$  $$Y_2:X_{2,\{3\}} \oplus X_{3,\{2\}},$$ $$Y_3: X_{1,\{3\}} \oplus X_{3,\{1\}},$$ $$Y_4: X_{1,\{2\}} \oplus X_{2,\{1\}} \oplus X_{2,\{3\}} \oplus X_{3,\{2\}},$$ $$Y_5: X_{1,\{2\}} \oplus X_{2,\{1\}} \oplus X_{1,\{3\}} \oplus X_{3,\{1\}} \text{ and }$$ $$Y_6: X_{2,\{3\}} \oplus X_{3,\{2\}} \oplus X_{1,\{3\}} \oplus X_{3,\{1\}}.$$
Decoding is done by syndrome decoding for error correcting index codes proposed in \cite{DSC}. It can be verified that the decoding is possible even if any one of the transmission goes in error. For instance, if the first transmission $Y_1$ is in error, from second and fourth transmissions, the first transmission is retrieved and the receivers are able to decode their demands. 
\end{exmp}

\begin{exmp}
	Consider the caching system in Example \ref{Ex:alpha2} with $N=K=3$, $r=M=2$. Assume that $N_e(\textbf{d})= 1$ and $\textbf{d}=(1,1,1)$. The side information hypergraph of the induced index coding problem is shown in Fig. \ref{fig2}(c).
	The transmission used by YMA scheme is $$X_{1,\{1,2\}} \oplus X_{1,\{2,3\}} \oplus X_{1,\{1,3\}}.$$ Suppose that we need to correct $\delta=2$ transmission errors, then from \cite{Gra} we have $N_2[1,5]=5$. Hence the optimal error correcting delivery scheme for this case is the concatenation of YMA scheme with $[5,1,5]_2$ length 5 binary repetition code.
		\[
		\bf{G}=
		\begin{bmatrix}
		1 & 1 & 1 & 1 & 1
		\end{bmatrix}.
		\]\\
		 Thus the optimal transmission scheme here is to repeat the YMA transmission five times.
\end{exmp}
\begin{exmp}
	Consider a caching system with $N=2$, $K=4$, $M=1$ and $r=2$. Let $N_e(\textbf{d})=2$ and let first and third user demand file $X_1$ and the second and fourth user demand file $X_2$. Thus $\textbf{d}=(1,2,1,2)$. The subfile division in symmetric batch prefetching is as follows:
	$$ X_1 : X_{1,\{1,2\}}, X_{1,\{2,3\}}, X_{1,\{1, 3\}}, X_{1,\{1,4\}}, X_{1,\{2,4\}}, X_{1,\{3,4\}}$$
	$$ X_2: X_{2,\{1,2\}}, X_{2,\{2,3\}}, X_{2,\{1,3\}}, X_{2,\{1,4\}}, X_{2,\{2,4\}}, X_{2,\{3,4\}}.$$
The cache placement according to symmetric batch prefetching is given by
$$ Z_1: X_{1,\{1,2\}},~X_{1,\{1,3\}},~X_{1,\{1,4\}}, ~X_{2,\{1,2\}},~X_{2,\{1,3\}}, ~X_{2,\{1,4\}} $$
$$ Z_2: X_{1,\{1,2\}},~X_{1,\{2,3\}},~X_{1,\{2,4\}}, ~X_{2,\{1,2\}},~X_{2,\{2,3\}}, ~X_{2,\{2,4\}} $$
$$ Z_3: X_{1,\{1,3\}},~X_{1,\{2,3\}},~X_{1,\{3,4\}}, ~X_{2,\{1,3\}},~X_{2,\{2,3\}}, ~X_{2,\{3,4\}} $$
$$ Z_4: X_{1,\{1,4\}},~X_{1,\{2,4\}},~X_{1,\{3,4\}}, ~X_{2,\{1,4\}},~X_{2,\{2,4\}}, ~X_{2,\{3,4\}}. $$
The side information hypergraph of the induced index coding problem is shown in Fig. \ref{fig3}. In this case $B= \{X_{1,\{2,3\}}, X_{1,\{2,4\}}, X_{1,\{3,4\}}, X_{2,\{3,4\}} \}.$

\begin{figure*}[!htbp]
	\centering
	\scalebox{0.8}{\includegraphics{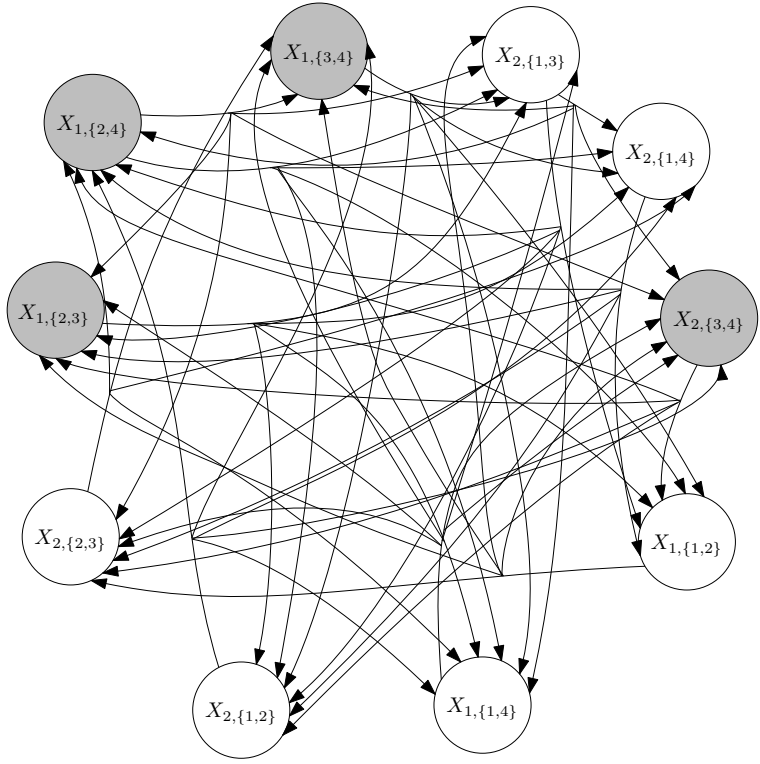}}
	\caption{Side information hypergraph of the induced index coding problem in Example \ref{ex:scheme1}.  The vertices in gray show the elements in $B$. } \label{fig3}
\end{figure*}
	
	The four transmissions in YMA scheme are
	$X_{1,\{2,3\}} \oplus X_{2,\{1,3\}} \oplus X_{1,\{1,2\}},~ 
	X_{1,\{2,4\}} \oplus X_{2,\{1,4\}} \oplus X_{2,\{1,2\}},~ 
	X_{2,\{3,4\}} \oplus X_{1,\{2,4\}} \oplus X_{2,\{2,3\}}$ and $ 
	X_{1,\{3,4\}} \oplus X_{1,\{1,4\}} \oplus X_{2,\{1,3\}}.$
	If we need to correct $\delta=1$ transmission error, then from \cite{Gra} we have $N_2[4,3]=7$. Therefore the optimal one error correcting delivery scheme here is the concatenation of YMA scheme with $[7,4,3]_2$ Hamming code. A generator matrix corresponding to this code is
		$$
		\bf{G}=
		\begin{bmatrix}
		1 & 0 & 0 & 0 & 0 & 1 & 1 \\
		0 & 1 & 0 & 0 & 1 & 0 & 1 \\
		0 & 0 & 1 & 0 & 1 & 1 & 0 \\
		0 & 0 & 0 & 1 & 1 & 1 & 1
		\end{bmatrix}.
		$$
	After concatenation, we get the following transmissions.
	$$Y_1: X_{1,\{2,3\}} \oplus X_{2,\{1,3\}} \oplus X_{1,\{1,2\}},$$ 
	$$Y_2: X_{1,\{2,4\}} \oplus X_{2,\{1,4\}} \oplus X_{2,\{1,2\}},$$ 
	$$Y_3: X_{2,\{3,4\}} \oplus X_{1,\{2,4\}} \oplus X_{2,\{2,3\}}, $$
	$$Y_4: X_{1,\{3,4\}} \oplus X_{1,\{1,4\}} \oplus X_{2,\{1,3\}},$$
	$$Y_5: Y_2 \oplus Y_3 \oplus Y_4, $$
	$$Y_6: Y_1 \oplus Y_3 \oplus Y_4 \text{ and }$$ $$ 
	Y_7 = Y_1 \oplus Y_2 \oplus Y_4.$$
	Decoding is possible even if one of the transmissions go in error. For example if $Y_1$ goes in error it can be retrieved by $Y_6 \oplus Y_3 \oplus Y_4$. 
	\label{ex:scheme1}
\end{exmp}
\begin{exmp}
	Consider a caching system with $N=K=N_e(\textbf{d})=4$, $r=M=1$. Let the $i$th user demands the file $X_i$. The subfile division in symmetric batch prefetching is as follows:
	$$ X_1 : X_{1,\{1\}}, X_{1,\{2\}}, X_{1,\{3\}}, X_{1,\{4\}},$$
	$$ X_2: X_{2,\{1\}}, X_{2,\{2\}}, X_{2,\{3\}}, X_{2,\{4\}},$$
	$$ X_3 : X_{3,\{1\}}, X_{3,\{2\}}, X_{3,\{3\}}, X_{3,\{4\}} \text{ and }$$
	$$ X_4 : X_{4,\{1\}}, X_{4,\{2\}}, X_{4,\{3\}}, X_{4,\{4\}}.$$
the cache contents of users according to symmetric batch prefetching are given as
$$ Z_1: X_{1,\{1\}}, X_{2,\{1\}}, X_{3,\{1\}}, X_{4,\{1\}}$$
$$ Z_2: X_{1,\{2\}}, X_{2,\{2\}}, X_{3,\{2\}}, X_{4,\{2\}}$$
$$ Z_3: X_{1,\{3\}}, X_{2,\{3\}}, X_{3,\{3\}}, X_{4,\{3\}}$$
$$ Z_4: X_{1,\{4\}}, X_{2,\{4\}}, X_{3,\{4\}}, X_{4,\{4\}}.$$
The side information hypergraph of the induced index coding problem is shown in Fig. \ref{fig4}. In this case $B= \{X_{1,\{2\}}, X_{1,\{3\}}, X_{1,\{4\}}, X_{2,\{3\}}, X_{2,\{4\}}, X_{3,\{4\}}  \}.$

\begin{figure*}[!htbp]
	\centering
	\scalebox{0.8}{\includegraphics{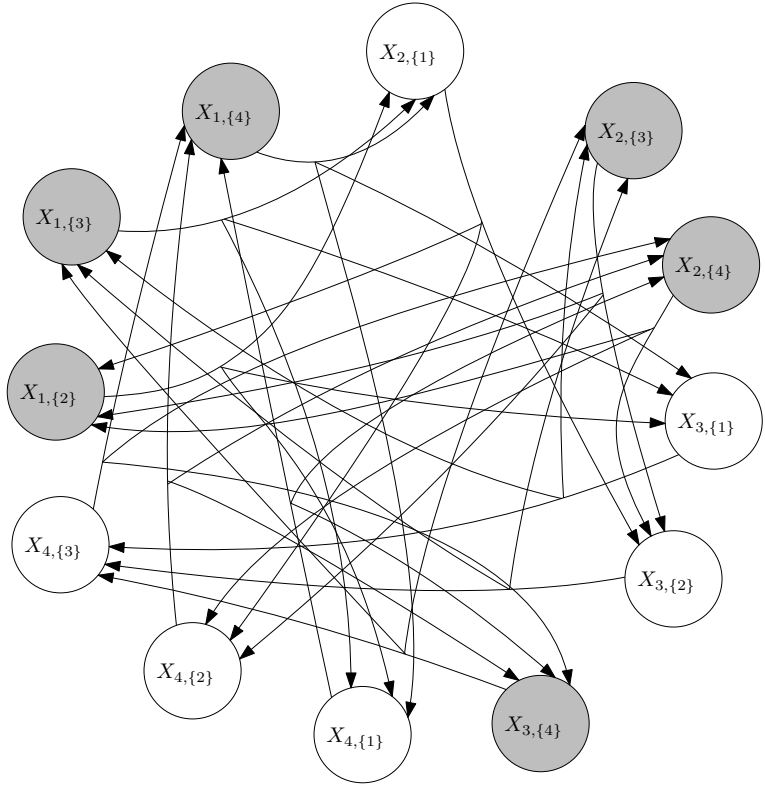}}
	\caption{Side information hypergraph of the induced index coding problem in Example \ref{ex:scheme2}.  The vertices in gray show the elements in $B$. } \label{fig4}
\end{figure*}

	The six transmissions in YMA scheme are
	$X_{1,\{2\}} \oplus X_{2,\{1\}},~ X_{1,\{3\}} \oplus X_{3,\{1\}},~X_{1,\{4\}} \oplus X_{4,\{1\}},~X_{2,\{3\}} \oplus X_{3,\{2\}},~X_{2,\{4\}} \oplus X_{4,\{2\}} \text{ and } X_{3,\{4\}} \oplus X_{4,\{3\}}.$
	If we need to correct $\delta=1$ transmission error, then from \cite{Gra} we have $N_2[6,3]=10$. Therefore the optimal one error correcting delivery scheme here is the concatenation of YMA scheme with $[10,4,3]_2$ code. A generator matrix corresponding to this code is
		$$
		\bf{G}=
		\begin{bmatrix}
		1 & 0 & 0 & 0 & 0 & 0 & 1 & 1 & 1 & 0 \\
		0 & 1 & 0 & 0 & 0 & 0 & 0 & 1 & 1 & 1 \\
		0 & 0 & 1 & 0 & 0 & 0 & 1 & 0 & 1 & 1
\\
		0 & 0 & 0 & 1 & 0 & 0 & 1 & 1 & 0 & 1
\\
		0 & 0 & 0 & 0 & 1 & 0 & 1 & 0 & 1 & 0 \\
		0 & 0 & 0 & 0 & 0 & 1 & 1 & 1 & 1 & 1
		\end{bmatrix}.
		$$
	After concatenation, we get the following transmissions.
	$$Y_1: X_{1,\{2\}} \oplus X_{2,\{1\}},$$ 
	$$Y_2: X_{1,\{3\}} \oplus X_{3,\{1\}},$$ 
	$$Y_3: X_{1,\{4\}} \oplus X_{4,\{1\}}, $$
	$$Y_4: X_{2,\{3\}} \oplus X_{3,\{2\}},$$
	$$Y_5: X_{2,\{4\}} \oplus X_{4,\{2\}}, $$
	$$Y_6: X_{3,\{4\}} \oplus X_{4,\{3\}},$$ 
	$$Y_7 = Y_1 \oplus Y_3 \oplus Y_4 \oplus Y_5 \oplus Y_6,$$
	$$Y_8 = Y_1 \oplus Y_2 \oplus Y_4  \oplus Y_6,$$
	$$Y_9 = Y_1 \oplus Y_2 \oplus Y_3 \oplus Y_5 \oplus Y_6 \text{ and }$$
	$$Y_{10} = Y_2 \oplus Y_3 \oplus Y_4 \oplus Y_6.$$
	Decoding is possible even if one of the transmissions go in error. Decoding is done by syndrome decoding for error correcting index codes proposed in \cite{DSC}. It can be seen that if $Y_1$ goes in error, it can be retrieved by $Y_6 \oplus Y_7 \oplus Y_8 \oplus Y_9$. 
	\label{ex:scheme2}
\end{exmp}

\label{sec:error_cor}

\section{Conclusion}
In this paper, we introduced the notion of error correcting coded caching scheme and characterized the minimum average rate and minimum peak rate for the problem. For coded caching schemes with symmetric batch prefetching, we found the closed form expressions for the average rate and peak rate. For these coded caching problems we proposed optimal error correcting delivery scheme. We also claimed the optimality of YMA scheme using  results from index coding.

\section*{Acknowledgment}
This work was supported partly by the Science and Engineering Research Board (SERB) of Department of Science and Technology (DST), Government of India, through J.C. Bose National Fellowship to B. Sundar Rajan.

\end{document}